\documentstyle[12pt]{article}

\begin{document}

\hfill WM-96-102

\hfill February 1996

 \medskip

\begin{center}
{\Large \bf Galactic Rotation Curves and
Linear\\[6pt] Potential Laws} \vglue 1cm

Carl E. Carlson
and
Eric J. Lowenstein\footnote{Present address: Department of
Mathematical Sciences, Clemson University,  Clemson,  SC 29634}

{\it Physics Department,
College of William and Mary,  Williamsburg,  VA  23187}


\end{center}

\begin{abstract}
We study the possibility that galactic rotation curves can be
explained by a gravitational potential that contains a
linear term as well as a Newtonian one.  This hypothesis,
suggested by conformal gravity, does allow good fits to the
galactic rotation curves of the galaxies we study, which
have a wide range of luminosities.  However, the
universality one might have expected of the parameter
describing the strength of the linear potential does not
emerge.  Instead, a different regularity is seen.
\end{abstract}

\section{Introduction}

The flat rotation curve seen in many
galaxies cannot be explained by the visible
matter and a Newtonian potential.   Often this
is taken as evidence for non-visible, i.e.,
``dark,'' matter in the galaxies.  There are
many candidates for what the dark matter could be (see for
example Griest (1995)).  Indeed, one of the remarkable
sucesses in the past year has been the observation of one
category of these candidates, the MACHOs or massive compact
halo objects by the EROS (Aubourg {\it et al.}\ 1993, 1995;
Beaulieu {\it et al.}\ 1995),  OGLE (Udalski {\it et al.}\
1993, 1994a,b,c), and MACHO (Alcock {\it et al.}\ 1993,
1995a,b,c; Bennett {\it et al.}\ 1994) collaborations. 
Unfortunately, these were discovered to be insufficiently
numerous to explain the flat rotation curves.

Whatever one's hopes or prejudices,
it is important to consider alternatives.  In the present
case, one alternative is that dark matter is not the cause
of the rotation curve problem, but rather that the
gravitational forces are not Newtonian at long distances. 
For example, Milgrom (1983, 1986, 1988, 1994) has suggested a
modified Newtonian dynamics wherein the gravitational
acceleration is the usually calculated Newtonian
gravitational acceleration if that acceleration is large
compared to some critical acceleration, but is the
geometrical mean of the critical acceleration and the
calculated Newtonian accleration if the Newtonian
acceleration is very small.  Fits to galactic data, after
some early doubts, seem possible using the same critical
acceration for each galaxy (Begeman, Broeils, \& Sanders
1991). Milgrom's idea gives an acceleration like what one
would get from a $1/r$ force, proportional to the square
root of the mass,  at very long range.

Here we shall examine a different alternative,
namely that, in addition to the Newtonian term, the potential
also contains a linear term.  This has been suggested on the
basis of conformal gravity (Mannheim \& Kazanas 1989, 1991;
Mannheim 1993), but one does not need to subscribe to this
viewpoint in order to evaluate the result. For a point mass
$M$, the potential (potential energy per unit
test mass) is

\begin{equation}
V = -{GM\over r} + \Gamma M r .
\end{equation}

\noindent The linear term must have a very small
coefficient so that it will contribute
noticeably only at very long distances.  The
coefficient of the linear potential,
$\Gamma$, should, like $G$, be a universal
coefficient if this potential correctly describes
nature.

We shall select a group of galaxies
for which a rotation curve has been well
measured, and for which sufficient other data
is available that can calculate a rotation
curve based on the directly observed matter. 
We shall, at least at the outset, allow ourselves to
vary only the mass to light ratio of the
luminous disk and the value of the linear
potential coefficient $\Gamma$.  The first goal
is to see if good fits to the rotation
curve can be obtained.  If the answer is
generally yes, then the further question will be
whether good fits can be obtained with the same
value of $\Gamma$ for each galaxy.  The answer
will be seen to be that good fits can be
obtained, but not with a universal value
of~$\Gamma$.  However, we will notice that a
fairly but not absolutely consistent value of
$\Gamma \times M_{galaxy}$ does emerge.  The latter is
equivalent to saying that when the gravitational forces get
very small, one does not get the Newtonian gravitational
acceleration but rather some small limiting value of
acceleration.  Finally, there will be some short discussion
of what is possible if some other parameters, such as the
distance to the galaxy, are allowed to vary.

\section{Testing the idea}

The galaxies we use are the ten used by Begeman {\it et
al.}\ (1991).
There are optical measurements giving
the luminosity, scale length, and distance of each of these
galaxies, and also radio measurements of the rotation curve
many scale lengths out from the center of the galaxy.  The
number of galaxies is restricted (Begeman {\it et al.}\ 1991) by
a requirement that they have reasonable azimuthal symmetry so
that the rotation curves accurately trace the overall mass
distribution of the galaxies.   The galaxies we use, as well as
their distances, luminosities, and scale lengths, are listed in
Table~\ref{table}.  The galaxies differ by a factor of 1000 in
luminosity and a factor 10 in size (scale length).  We will
not, at least for now, vary the distances and scale lengths
found in the literature.

We give a few details of how we proceed.

The visible mass of each galaxy includes luminous matter
and H{\sc i} gas.  The luminosity area density of the
galactic disk is generally well represented by a
falloff that is exponential in distance from
the galactic center.  The scale lengths $R_D$ are given in
Table~\ref{table}.  We take the mass to light ratio for
the luminous matter in the disk,
$M/L$, to be constant within a given galaxy although we
allow it to vary from galaxy to galaxy.  The mass area
density of  the luminous disk, $\sigma_D$, is then

\begin{equation}
\sigma_D  
  = \left( M_D \over 2 \pi R_D^2 \right)
                                   e^{-r/R_D}.
\end{equation}

\noindent where $M_D$ is the mass of the luminous matter
in the disk. 

The point mass Newtonian plus linear potential needs
to be integrated over the expontential disk mass
distribution to get the potential for a galaxy.  Mannheim
(1993, 1995) has given the result extending earlier work by
others for the Newtonian case.  The Newtonian force per unit
test mass (acceleration) is
 
\begin{equation}
g_{ND} = {G M_D \over R_D^2} \alpha
     \left[ I_0(\alpha) K_0(\alpha) 
        -   I_1(\alpha) K_1(\alpha) \right]
\end{equation}

\noindent where

\begin{equation}
\alpha \equiv r / 2 R_0  .
\end{equation}

\noindent and $I_\nu$ and $K_\nu$ are Bessel functions. 
The corresponding result for the linear potential is

\begin{equation}
g_{LD} = 2 \Gamma M_D 
      \alpha I_1(\alpha) K_1(\alpha)   .
\end{equation}

Two of the galaxies, the two biggest, in our sample have
central bulges which are not well included by a single
exponential falloff.  For these galaxies the luminosity
curve is fit with a sum of two exponentials.  The bulge
and (main) disk components are allowed different $M/L$
ratios, and in Table~\ref{table} we list the scale
lengths, fitted $M/L$ ratios, and total mass of the two
components separately.

In addition, the H{\sc i} gas in the galaxies is visible
through its 21 cm radiation. If the distance to the
galaxy is correctly known, the measurements give the mass
of the H{\sc i} directly, and we increase this mass by a
factor 4/3 to account for heavier material, mainly
primordial He.  The gas contribution is dynamically
significant only for the three lightest galaxies in our
sample, and for each of these the gas mass area density
is well fit by a gaussian,

\begin{equation}
\sigma_G = {M_G \over \pi R_G^2 } e^{-r^2/R_G^2}  ,
\end{equation}

\noindent  and the total gas mass $M_G$ and gaussian
scale length for the gas $R_G$ are listed
in Table~\ref{table2}.  We here record the acceleration fields
due to the gas and the Newtonian and linear potentials
(Mannheim 1995):

\begin{equation}
g_{NG} = {G M_G r \over R_G^3 } \sqrt{\pi} e^{-\beta}
         \left( I_0(\beta) - I_1(\beta) \right)
\end{equation}

\noindent where

\begin{equation}
\beta \equiv r^2/2 R_G^2
\end{equation}

\noindent and

\begin{equation}
g_{LG} = {\Gamma M_G r \over 2 R_G} \sqrt{\pi} e^{-\beta}
         \left( I_0(\beta) + I_1(\beta) \right).
\end{equation}

Thus in a case where we include gas and a luminous disk
described by a single exponential we have

\begin{equation}
g(r) = {v^2(r) \over r} = g_{ND}+g_{NG}+g_{LD}+g_{LG},
\end{equation}

\noindent where $v(r)$ is the revolution velocity of the
galaxy.

We then fit to the galactic rotation curves,  given the
distribution and luminosity of the luminous matter and the
distribution and actual mass of the gas.  To repeat, the
unknown quantities for each galaxy are $M/L$ for the
luminous matter and $\Gamma$. The rise of the rotation curve
to its peak is mainly determined by the Newtonian part of
the gravitational force (the contribution from the linear
potential is numerically small in this region),  and this
rise determines
$M/L$.  Conversely, the contribution of the Newtonian
potential to the far out part of the rotation curve is small,
and this part of the curve basically determines $\Gamma$ for
each galaxy.

Our fits to the rotation curves are shown in
Fig.~\ref{rotationcurves}.  One sees that the
rotation curve fits, based on two parameters (three for
NGC 2841 and 7331), are tolerably good.  The curves in
Fig.~\ref{rotationcurves} also show a clear feature of a
linear plus Newtonian potential in that the observed
near-flatness depends upon interplay between the two
contributions, and if the rotation curve were measured
farther out the curve would rise.

The values of $M/L$ and $\Gamma$ we get for each galaxy
are shown in Table~\ref{table}.  Although we have good
fits, the price is that the largest and smallest values of
$\Gamma$ in the Table differ by two orders of magnitude. 
We can get an idea of the play in
$\Gamma$ by asking what values we get if we set $M/L$ to
zero (which gives maximum possible $\Gamma$ at the expense
of a poor small radius fit), or set $M/L$ to twice its
best value (which also gives a poor small radius fit). 
One gets changes of about $\pm 5\%$ in $\Gamma$ for the gas
dominated galaxy DDO 154 to typically $\pm 30\%$ for a
galaxy where the known gas plays little dynamical
r\^ole.  Thus the fits cannot be modified to get the same
$\Gamma$ for each galaxy and we conclude that the values of
$\Gamma$ are not universal.

\subsection{Allowing the distance to vary}

Distances to galaxies are of course not perfectly measured.
Indeed, for DDO 154 there is discussion (Carignan \& Beaulieu
1989) of whether it is part of the Canes Venatici I cluster at
4 Mpc or really part of the Coma I cluster beyond it at 10
Mpc.  So we may consider how changes in the measured distance
will affect the values of $\Gamma$ and $M_{gal}\Gamma$.  For
galaxies dominated by luminous matter, when the distance
scales like
$d\rightarrow \eta d$, then $r \rightarrow \eta r$ and
$L\rightarrow \eta^2 L$.  Then choosing 
$(M/L)\rightarrow \eta^{-1} (M/L)$ leads to unchanged
rotation curves provided

\begin{equation}
\Gamma_{new} = \eta^{-2} \Gamma_{old} =
    \left(d_{old}\over d_{new}\right)^2 \Gamma_{old}   .
\end{equation}

For gas dominated galaxies, we have directly 
$M\rightarrow \eta^2 M$ and examining the far
out part of the rotation curve then leads to

\begin{equation}
\Gamma_{new} = 
   \left(d_{old}\over d_{new} \right)^3 \Gamma_{old}  .
\end{equation}

For either case,

\begin{equation}
\left( M_{gal} \Gamma \right)_{new} = 
       {d_{old}\over d_{new}} 
               \left( M_{gal} \Gamma \right)_{old}  .
\end{equation}

To reconcile or make the same all the values of $\Gamma$ would
involve moving galactic distances so that the smaller galaxies
were systematically moved out  by a factor of 5 to 10 compared
to the larger.  We will not entertain this idea.  Reconciling
$M_{gal} \Gamma$ is less motivated theoretically, but it is
relatively easy to do.  Choosing to set $M_{gal}
\Gamma = 2.42$ (the geometric mean of the relevant numbers in
Table~\ref{table}), a two parameter fit varying $d$ and $M/L$
produces rotation curves like the ones we have already shown,
with the scaling of $M_{gal}\Gamma$ working as suggested
above even for the cases where luminous matter and gas are
both important.  Reconciling $M_{gal}\Gamma$ in this way
requires, in the extreme cases, having DDO 154 at 2.4 Mpc
instead of 4 Mpc and NGC 2841 at 22.5 Mpc instead of 9.46
Mpc.  In fact, for NGC 2841 the distance is already in dispute
(Sanders \& Begeman 1994) since the distance derived from the
Tully-Fisher relation is about twice 9.46 Mpc obtained from
Hubble's law and used here.  Sanders \& Begeman (1994) suggest
2841's recession speed is greatly affected by proximity to the
Virgo Cluster, and that the larger distance is more likely
correct.

\section{Conclusion}

We have attempted to confirm or disconfirm the idea that the
far out part of the galactic rotation curves, usually taken
as evidence of dark matter, may be well described by a linear
add-on to the Newtonian potential.  The original suggestion
was that the coefficient of the linear potential be a
universal constant $\Gamma$ times the mass of the source,
just as the coefficient in the Newtonian potential is a
universal constant $G$ times the mass of the source.  This is
what one would expect if gravity theory, even with
modifications, were a metric theory driven by the
energy-momentum tensor of the source.  

The original suggestion works poorly.  Decent fits can be
gotten for the rotation curves of a selection of smooth and
azimuthally symmetric galaxies, but the fits require very
different values of $\Gamma$ for different size galaxies.

The galactic data, despite lacking {\it a priori} theoretical
motivation, do seem to give $M_{gal}\times\Gamma$ nearly the
same for each galaxy.  In other words, the centripetal
acceleration of matter near the galactic edges approaches a
limiting value which is about the same for any galaxy. 
Numerically, the value $g_0 = M_{gal}\Gamma$ is about
$2\frac{1}{2}\times10^{-11} m/s^2$.  Thus there is some
systematic and reproducible feature of the mysterious
galactic rotation curves.  Such things have also been noted in
the context of the dark matter
explanation (Bahcall and Casertano 1985; van Albada \&
Sancisi 1986), and may be a clue to an underlying
understanding of galactic structure or binding.

\section*{Acknowledgment}

Mannheim and Kmetko (1996) have studied the galactic rotation
curves from the same viewpoint, and have come to the same
conclusions.  We thank Philip Mannheim for much friendly
communication while this work was progressing.  CEC thanks the
National Science Foundation for support under Grant
PHY-9306141.

\newpage

\begin{table}[h]
\hglue -0.35in 
\begin{tabular}{lccccccc}\hline
\   & $d$ & 
    $L$  & 
    $R_D$   &
    $M/L $ &
    $ \Gamma  $   &
    $ M_{gal} $  &
    $ \Gamma M_{gal} $ \rule{0pt}{13pt}\\
Galaxy  & Mpc & $10^9 L_\odot$ & kpc & $M_\odot/L_\odot$
       & $10^{-52}N/kg^2$ & $10^9 M_\odot$ 
       & $10^{-11} m/s^2$ \rule{0pt}{13pt} \\[2pt]
\hline
DDO 154  & \ 4.00 & \ 0.05 & 0.50 & 1.2 & 164 &0.42 & 1.37 \\
DDO 170  &  12.01 & \ 0.16 & 1.28 & 4.5 &  46 & 1.6 & 1.46 \\
NGC 1560 & \ 3.00 & \ 0.35 & 1.30 & 2.3 &  60 & 1.9 & 2.27 \\
UGC 2259 & \ 9.80 & \ 1.02 & 1.33 & 4.1 &  29 & 4.2 & 2.38 \\
NGC 6503 & \ 5.94 & \ 4.80 & 1.73 & 3.0 & 5.8 & 14  & 1.67 \\
NGC 2403 & \ 3.25 & \ 7.90 & 2.05 & 2.3 & 6.9 & 18  & 2.49 \\
NGC 3198 & \ 9.36 & \ 9.00 & 2.63 & 3.8 & 3.1 & 34  & 2.14 \\
NGC 2903 & \ 6.40 &  15.30 & 2.02 & 3.6 & 3.0 & 55  & 3.31 \\
NGC 2841 & \ 9.46 &  20.50 & 0.50 & 3.5 & 1.9 & 150 & 5.75 \\
         &      &(5.9+14.6)& 2.38 & 9.0 &  & &             \\ 
NGC 7331 &  14.90 &  54.00 & 1.20 & 0.75& 1.3 & 140 & 3.71 \\
         &     &(31.5+22.5)& 4.48 & 5.2 &&&                \\
\hline
\end{tabular}

\caption{The galaxies.  The two brightest galaxies have
two component expontentials describing their luminosity
profiles.  Both scale lengths are given, and the
luminosities of the inner (``bulge'') and outer parts of the
disk are given, in that order, parenthetically in the
luminosity column.}
\label{table}

\end{table}

\begin{table}

\centering
\begin{tabular}{cccc}\hline
Galaxy & $M_{\rm H{\scriptscriptstyle I}} (M_\odot)$ 
     & $M_G (M_\odot)$ & $R_G$
\\
\hline
DDO 154 & $2.7 \times 10^8$ & $3.6 \times 10^8$ 
    &$3.3'=3.8 {\rm kpc}$\\ 
DDO 170 & $6.6 \times 10^8$ & $8.8 \times 10^8$ 
    & $95''=6.7 {\rm kpc}$\\
NGC 1560 & $8.2 \times 10^8$ & $10.9 \times 10^8$ 
    & $5.6'=4.85 {\rm kpc}$ \\
\hline
\end{tabular}
\caption{Parameters for gas in three galaxies.}

\label{table2}

\end{table}

\newpage

\section*{References}

\parindent 0 pt

Alcock, C., {\it et al.}\ 1993, Nature, 365, 621
    
------. 1995a, Phys. Rev. Lett., 74, 2867

------. 1995b, ApJ, 445, 133

------. 1995c, ApJ, 449, 28

Aubourg, E. {\it et al.}\ 1993, Nature, 365, 623

------. 1995, A \& A, 301, 1
        
Bahcall, J. N. \& Casertano, S. 1985  ApJ, 293, L7

Beaulieu J.P., {\it et al.}\ 1995, A \& A, 299, 168

Begeman, K. G.,  Broeils, A. H., \& Sanders, R. H. 1991,
     MNRAS, 249, 523

\parindent 20 pt \hang \noindent
Bennett, D.P., {\it et al.} 1994, Proceedings of the 5th
        Astrophysics Conference in Maryland: Dark Matter
        \parindent 0 pt

Carignan, C. \& Beaulieu, S. 1989, ApJ, 347, 760

\parindent 20 pt \hang \noindent
Griest, K., Lectures presented at
   the International School of Physics ``Enrico Fermi" Course
   ``Dark Matter in the Universe", Varenna, 25 July - 4 August,
   1995  \parindent 0 pt
    
Mannheim, P. D. \& Kazanas, D. 1989, ApJ, 342, 635

------. 1991, Phys. Rev. D, 44, 417

Mannheim, P. D. 1993, ApJ, 419, 150

------. 1995, preprint UCONN 95-02; astro-ph/9504022

Mannheim, P. D., \& Kmetko, J. 1996, preprint UCONN 96-02

Milgrom, M. 1983, ApJ, 270, 365
    
------. 1986, ApJ, 302, 617
     
------. 1988, ApJ, 333, 689

------. 1994, Ann. Physics (N. Y.), 229, 384

Sanders, R. H., \& Begeman, K. G. 1994, MNRAS, 266, 360

Udalski, A., {\it et al.}\ 1993, Acta Astronomica, 43, 289
    
------. 1994a, Acta Astronomica, 44, 165
    
------. 1994b, Acta Astronomica, 44, 227
    
------. 1994c, ApJ Lett., 436, L103.

\parindent 20 pt \hang \noindent
van Albada, T. S. \& Sancisi, R. 1986,
     Phil.\ Trans.\ R. Soc.\ Lond.\ A 320, 447

\newpage

\begin{figure}


\caption{\protect\small {\it Caption:} Rotation curves.  In all
cases, $v$ is in km/sec and $r$ is in kpc.  The heavy line is
the full result. The dotted line would be the result from the
linear potential alone, the solid line would be the result
from the Newtonian potential alone.  In the cases where there
is a two component fit to the mass distribution, we have
indicated the separate contributions to the Newtonian result
with dashed lines.  For the lighter galaxies, the two
components are luminous matter and gas, with the gas
contribution peaking farther out; for the heavier galaxies,
both components represent luminous matter.}
\label{rotationcurves}

\end{figure}

\end{document}